\documentclass[proceedings]{JHEP3}
\PrHEP{PrHEP unesp2002} 

\def\nonu{\nonumber}

\def\br{\begin{eqnarray}}
\def\er{\end{eqnarray}}
\def\be{\begin{equation}}
\def\ee{\end{equation}}
\def\0{\nonumber}

\def\({\left(}
\def\){\right)}

\relax

\def\a{\alpha}

\def\d{\delta}

\def\eps{\epsilon}

\def\l{\lambda}

\def\o{\over}

\def\pa{\partial}
\def\pr{\prime}

\def\lie{{\cal G}}

\def\rlx{\relax\leavevmode}
\def\inbar{\vrule height1.5ex width.4pt depth0pt}
\def\IZ{\rlx\hbox{\sf Z\kern-.4em Z}}
\def\IR{\rlx\hbox{\rm I\kern-.18em R}}
\def\IC{\rlx\hbox{\,$\inbar\kern-.3em{\rm C}$}}
\def\one{\hbox{{1}\kern-.25em\hbox{l}}}

\def\NPB#1#2#3{{\sl Nucl. Phys.} {\bf B#1} (#2) #3}

\def\PLB#1#2#3{{\sl Phys. Lett.} {\bf #1B} (#2) #3}

\def\AoP#1#2#3{{\sl Ann. of Phys.} {\bf #1} (#2) #3}

\def\MPLA#1#2#3{{\sl Mod. Phys. Lett.} {\bf A#1} (#2) #3}

\conference{Workshop on integrable theories, solitons and duality.}
\title{Axial Vector Duality in  Affine NA  Toda Models}
\author{{ J.F. Gomes} \speaker{
 { G.M. Sotkov}} and { A.H. Zimerman}\\
 Instituto de F\'\i sica Te\'orica - IFT/UNESP,\\ 
Rua Pamplona 145\\
01405-900, S\~ao Paulo - SP, Brazil\\
E-mail \email{sotkov@ift.unesp.br} }
\abstract{A general and systematic  construction of Non Abelian affine Toda models and its symmetries 
is proposed in terms of its underlying Lie algebraic structure.   
It is also shown that such class of two dimensional integrable models naturally 
leads to the construction of a pair of
actions related by T-duality transformations.}

\begin{document}

\section{Introduction}

Two dimensional integrable models represent  an important laboratory
for testing new ideas and developing new methods for constructing exact
solutions as well as for the nonperturbative quantization of 4-D non-abelian
gauge theories, gravity and  string theory.  Among the numerous
techniques for constructing 2-D integrable models   and their solutions
\cite{fadeev}, \cite{lez-sav}, the hamiltonian reduction of the Wess-Zumino-Witten (WZW) model 
(or equivalently the gauged WZW ) 
associated to a finite dimensional Lie algebra $\lie $ has provided an universal 
and simple method for deriving the equations of motion
(or action ) of 2-d integrable models.  In particular, the conformal Toda (CT) 
models were constructed by implementing a consistent set of
constraints on the WZW currents  \cite{ora}.  The method was subsequently extended to
construct the conformal affine Toda models (CAT) from 
infinite dimensional affine algebras,
leading to WZW currents satisfying the so called two loop current algebra \cite{Aratyn} \cite{schw}.

The affine Toda models consists of a class of relativistic  two dimensional 
integrable models admiting soliton solutions with non
trivial topological charge (e.g. the abelian affine Toda models).
Among such models we encounter the  Non Abelian affine (NA) Toda models 
which  admit electrically charged solitons \cite{tau}.  In general,  the NA Toda models admit  solitons
with non trivial internal symmetry structure.  
 The formulation and classification of such  models with its global symmetry structure is given 
in terms of the decomposition of an underlying Lie algebraic structure 
according to a grading operator $Q$ and, in terms of a pair of constant generators 
$\eps_{\pm}$ of grade $\pm 1$. In particular, integrable perturbations 
of the WZW characterized by $\eps_{\pm}$ describe the
dynamics of the   fields  parametrizing the zero grade subalgebra $\lie_0$. 
The action manifests  chiral symmetry associated to the subalgebra 
$\lie_0^0 \subset \lie_0$ due to the fact that $Y \in \lie_0^0, [Y, \eps_{\pm}]=0$.
The existence of such  subalgebra allows the implementation of 
subsidiary constraints within $\lie_0^0$ and   the reduction of  the model from the 
group $G_0$ to the coset $G_0/G_0^0$. The structure of the coset $G_0/G_0^0$  
viewed according to axial or vector
gauging  leads to different parametrizations and different actions.

We first discuss the general construction of NA Toda models 
and its different internal symmetry structure. Next, we discuss the 
structure of the coset $G_0/G_0^0 = SL(2)\otimes U(1)^{n-1}/ U(1)$ 
according to axial and vector
gaugings and explicitly construct the associated lagrangians. Subsequently, we show how the 
chiral symmetry   of the group model becomes  global under the reduction to the  
coset. Finally, we show that 
the axial and vector models are related by  canonical transformation 
\cite{dual} preserving the Hamiltonian which also   transforms 
 topological into  electric current and vice-versa.

\section{General Construction of Toda Models}

The basic ingredient in constructing Toda models is the decomposition of a 
Lie algebra $\lie $ of finite  or infinite dimension in
terms of graded subspaces defined according to a grading operator $Q$, 
\br
\quad [ Q, \lie_l] = l \lie_l, \quad \quad 
\lie =\oplus \lie_l, \quad \quad [\lie_l, \lie_k ] \subset \lie_{l+k}, \; l,k= 0, \pm 1, \cdots 
\label{2.1}
\er
In particular, the zero grade subspace $\lie_0$ plays an important 
role since it is parametrized by the Toda fields.
The grading operator $Q$ induces the notion of negative and positive grade subalgebras
and henceforth the decomposition of a group element in the Gauss form, 
\br
g=NBM
\label{2.3}
\er
where $N=\exp {(\lie_{<})}$, $B=\exp {(\lie_{0})}$ and $M=\exp {(\lie_{>})}$.

The action for the Toda fields is constructed from the gauged Wess-Zumino-Witten (WZW)
 action \cite{tau}, \cite{elek}, 
\br
S_{G/H}(g,A,\bar{A})&=&S_{WZW}(g) \nonumber \\
&-&\frac{k}{2\pi}\int d^2x Tr\( A(\bar{\partial}gg^{-1}-\epsilon_{+})
+\bar{A}(g^{-1}\partial g-\epsilon_{-})+Ag\bar{A}g^{-1}  \) 
\label{2.4}
\er
where $A = A_- \in \lie_{<}, \; \bar A = \bar A_+ \in \lie_{>}$, $\eps_{\pm}$ are 
constant operators of grade $\pm 1$.  The action (\ref{2.4}) is invariant under
\br
 g^{\prime}=\alpha_{-}g\alpha_{+}, \quad 
 A^{\prime}=\alpha_{-}A\alpha_{-}^{-1}
+\alpha_{-}\partial \alpha_{-}^{-1},
\quad  
 \bar{A}^{\prime}=\alpha_{+}^{-1}\bar{A}\alpha_{+}
+\bar{\partial}\alpha_{+}^{-1}\alpha_{+},  
\label{2.5} 
\er
 where $\a_{-}  \in \lie_{<}, \; \a_{+}  \in \lie_{>}$.
It therefore follows that  $ S_{G/H}(g,A,\bar{A}) =S_{G/H}(B,A^{\pr},\bar{A}^{\pr}) $.
 
Integrating over the auxiliary fields $A, \bar A$, we find the effective action,
\br
S_{eff}(B) =  S_{WZW}(B)- 
 {{k\o {2\pi}}} \int Tr \( \eps_+ B  \eps_- B^{-1}\) d^2x   
  \label{2.6}
\er
The equations of motion are given by
\br 
\bar \pa (B^{-1} \pa B) + [ {\eps_-}, B^{-1}  {\eps_+} B] =0, \quad  
\quad \pa (\bar \pa B B^{-1} ) - [ {\eps_+}, B {\eps_-} B^{-1}] =0
\label{2.7}
\er
It is straightforward to derive from the eqns. of motion (\ref{2.7}) that 
 chiral currents are associated to the subalgebra $\lie_0^0
\subset \lie_0$ defined as 
$\lie_0^0 = \{ X \in \lie_0, \;\; {\rm such \;\; that } \;\; [ X, \eps_{\pm} ] =0 \}$,
 i.e.,
\br
J_X = Tr \( X B^{-1} \pa B \), \quad \quad \bar J_X = Tr \( X \bar \pa B B^{-1}\), 
\quad \quad \bar \pa  J_X = \pa \bar J_X =0
\label{2.7a}
\er
In order to illustrate the different algebraic structure,  consider 
the following examples within the affine $\lie = \hat {SL}(n+1)$:
\begin{enumerate}
\item {\it Abelian affine Toda}  
\br
Q&=& (n+1)d + \sum_{l=1}^{n} \l_l\cdot H, \nonu \\
 \lie_0 &=& U(1)^n = \{h_1, \cdots , h_n \} \nonu \\
\eps_{\pm} &=& \sum_{l=1}^{n} E_{\pm \a_l}^{(0)} + E_{\mp (\a_1 + \cdots + \a_n)}^{(\pm 1)} \nonu \\
\lie_0^0 &=& \emptyset
\label{2.8}
\er
\item {\it Non Abelian  affine Toda}  
\begin{enumerate}
\item
\br
Q&=& n d + \sum_{l=2}^{n} \l_l\cdot H, \nonu \\
\lie_0 &=& SL(2)\otimes U(1)^{n-1} = \{E_{\pm \a_1}, h_1, \cdots , h_n \} \nonu \\
\eps_{\pm} &=& \sum_{l=2}^{n} E_{\pm \a_l}^{(0)} + E_{\mp (\a_2 + \cdots + \a_n)}^{(\pm 1)}\nonu \\
\lie_0^0 &=& U(1) = \{ \l_1 \cdot H \}
\label{2.9}
\er
\item 
\br
Q&=& (n-1) d + \sum_{l=2}^{n-1} \l_l\cdot H, \nonu \\
\lie_0 &=& SL(2)\otimes SL(2)\otimes U(1)^{n-2} = \{E_{\pm \a_1}, E_{\pm \a_n}, h_1, \cdots , h_n \} \nonu \\
\eps_{\pm} &=& \sum_{l=2}^{n-1} E_{\pm \a_l}^{(0)} + E_{\mp (\a_2 + \cdots + \a_{n-1})}^{(\pm 1)} \nonu \\
\lie_0^0 &=& U(1)\otimes U(1) = \{ \l_1 \cdot H, \l_n \cdot H \}
\label{2.10}
\er
\item 
\br
Q&=& (n-1) d + \sum_{l=3}^{n} \l_l\cdot H, \nonu \\
\lie_0 &=&  SL(3)\otimes U(1)^{n-2} = \{E_{\pm \a_1}, E_{\pm \a_2},  E_{\pm (\a_1+\a_2)}, h_1, \cdots , h_n \} \nonu \\
\eps_{\pm} &=& \sum_{l=3}^{n} E_{\pm \a_l}^{(0)} + E_{\mp (\a_3 + \cdots + \a_{n})}^{(\pm 1)}\nonu \\
\lie_0^0 &=& SL(2)\otimes U(1) = \{ E_{\pm \a_1}, \l_1 \cdot H, \l_n \cdot H \}
\label{2.11}
\er
\end{enumerate}
\end{enumerate}
For the cases where $\lie_0^0 \neq 0$, we may impose consistently 
the additional constraints $J_{X} = \bar J_{X} = 0, X \in \lie_0^0$. 
The construction of the gauged WZW action  taking into account the
subsidiary constraints reduces the model from the group $G_0$ to 
the coset $G_0/G_0^0$ and is given by
\br
S_{G_0/G_0^0}(B,{A}_{0},\bar{A}_{0} ) &=&  S_{WZW}(B)- 
 {{k\o {2\pi}}} \int Tr \( \eps_+ B  \eps_- B^{-1}\) d^2x\nonu \\   
  &-&{{k\o {2\pi}}}\int Tr\( \pm  A_{0}\bar{\partial}B
B^{-1} + \bar{A}_{0}B^{-1}\partial B
\pm  A_{0}B\bar{A}_{0}B^{-1} + A_{0}\bar{A}_{0} \)d^2x \nonu \\
\label{2.12}
\er
where the $\pm $ signs correspond to axial or vector gaugings respectively.
The action (\ref{2.12}) is invariant under
\br
 B^{\pr} = \a_0 B \a_0^{\pr}, \quad \quad 
A_0^{\pr} = A_0 - \a_0^{-1} \pa \a_0, \quad \quad \bar A_0^{\pr} = \bar A_0 - \bar  \pa \a_0^{\pr}(\a_0^{\pr})^{-1}
\label{2.13}
\er
 and $\a_0^{\pr} =
\a_0(z, \bar z) \in \lie_0^0$ for axial and $\a_0^{\pr} =
\a_0^{-1}(z, \bar z) \in \lie_0^0$ for vector cases, i.e.,
\br
S_{G_0/G_0^0}(B,{A}_{0},\bar{A}_{0} ) =  S_{G_0/G_0^0}(\a_0 B \a_0^{\pr} = g_0^f,{A^{\pr}}_{0},\bar{A^{\pr}}_{0} )  
\label{2.14}
\er

\section{The structure of the coset $G_0/G_0^0$}

In this section we discuss the structure of the coset $G_0/G_0^0$  constructed  
according to axial and vector gaugings.  We shall be considering  the NA Toda 
models of case $(2a)$ where $\lie_0^0 = U(1)$.  The group element of 
the zero grade subgroup $G_0$ is parametrized as 
\br
B = e^{\tilde \chi E_{-\a_1}} e^{R\l_1 \cdot H + \sum_{l=2}^{n}\varphi_l h_l}e^{\tilde \psi E_{\a_1}}
\label{3.1}
\er
According to the axial gauging we can write $B$ as an element of the
the zero grade subgroup $G_0$ is parametrized as 
\br
B = e^{{1\o 2}R\l_1 \cdot H}\( g_{0, ax}^f \) e^{{1\o 2}R\l_1 \cdot H}, \quad \quad {\rm where} \quad \quad
g_{0, ax}^f =  e^{\tilde \chi e^{{1\o 2}R}E_{-\a_1}} 
e^{ \sum_{l=2}^{n}\varphi_l h_l}e^{ \tilde \psi e^{{1\o 2}R} E_{\a_1}}
\label{3.2}
\er
The effective action is obtained  integrating (\ref{2.12}) 
over $A_0, \bar A_0$,  yielding \cite{tau}
\br
{\cal L}_{eff}^{ax}={1\o 2} \sum_{a,b =2}^{n}\eta_{ab} \pa \varphi_a \bar \pa \varphi_b  + 
{1\o 2} {{\bar \pa \psi \pa \chi }\o \Delta
 }e^{-\varphi_2}  -  V_{ax}, \quad \Delta = 1 + {{n+1}\o {2n}}\psi \chi
e^{-\varphi_2}
 \label{3.3}
 \er
 where $ \psi = \tilde \psi e^{{1\o 2}R}, \chi = \tilde \chi e^{{1\o 2}R}$, and 
  $V_{ax} = \sum_{l=2}^n e^{2\varphi_l - \varphi_{l-1}- \varphi_{l+1}} + 
 e^{\varphi_2+\varphi_n} ( 1+ \psi \chi
 e^{-\varphi_2})$.
 
The vector gauging can be implemented from the zero grade subgroup $G_0$ written as
\br
B = e^{u\l_1 \cdot H}\( g_{0, vec}^f \) e^{-u\l_1 \cdot H}, \quad \quad {\rm where} \quad \quad
g_{0, vec}^f =  e^{\tilde \chi e^{u}E_{-\a_1}} 
e^{  \sum_{l=1}^{n}\phi_l h_l}e^{ \tilde \psi e^{-u} E_{\a_1}}
\label{3.3a}
\er
Since $u$ is arbitrary, we may choose $u = {1\o 2} ln \( -{{\tilde \psi }\o {\tilde \chi }}\) $ so that
\br
g_{0, vec}^f =  e^{-t E_{-\a_1}} 
e^{ \sum_{l=1}^{n}\phi_l h_l}e^{t E_{\a_1}}, \quad \quad t^2 = -\tilde \psi \tilde \chi 
\label{3.4}
\er
The effective action for the vector model is \cite{dual}   
\br
{\cal L}_{eff}^{vec}&=&{1\o 2} \sum_{a,b =1}^{n}\eta_{ab} \pa \phi_a \bar \pa \phi_b   
+ {{\pa \phi_1 \bar \pa \phi_1 }\o {t^2}}e^{-2\phi_1+\phi_2 } \nonu \\
&+& \pa \phi_1 \bar \pa ln (t) + \bar \pa \phi_1 \pa ln (t) - V_{vec}.
 \label{3.5}
 \er
In terms of a more convenient set of variables,
\br
A= e^{\phi_1} = e^{{{n}\o {n+1}}R}, \quad B = A^{-1}(1-t^2 e^{2\phi_1-\phi_2}), 
\quad C_i = e^{\phi_{i+2}-\phi_{i+1}- {{R}\o n}}
\label{3.6}
\er
we get
\br
{\cal L}_{eff}^{vec}&=&{1\o 2} \sum_{l=1}^{n-1} \( \pa ln (C_i) \bar \pa ln (C_iC_{i+1} \cdots C_{n-1}) +
 \bar \pa ln (C_i)  \pa ln (C_iC_{i+1} \cdots C_{n-1})  \)\nonu \\
 &-& {1\o 2} {{\pa A \bar \pa B + \bar \pa A \pa B}\o {1-AB}} -V_{vec}
\label{3.7}
 \er
where $V_{vec} = AC_1^2C_2 \cdots C_{n-1} + {{C_2}\o {C_1}}+ {{C_3}\o {C_2}}+ 
\cdots {{C_{n-2}}\o {C_{n-1}}}+ {{B}\o {C_1C_2 \cdots C_{n-1}^2}}$.

\section{Chiral and Global Symmetries, from  $G_0$ to  $G_0/G_0^0$}

Consider the unconstrained model defined in the group $G_0$ with action
\br
S_{eff}(B) =  S_{WZW}(B)- 
 {{k\o {2\pi}}} \int Tr \( \eps_+ B  \eps_- B^{-1}\) d^2x   
\label{4.1}
\er
The action (\ref{4.1}) is invariant under chiral transformation
\br
B^{\pr}(\bar z, z) = e^{\bar w (\bar z)\l_1 \cdot H} B(\bar z, z) e^{ w(z)\l_1 \cdot H}, 
\label{4.2}
\er
which in components reads 
\br
\tilde \chi^{\pr} = \tilde \chi e^{\bar w (\bar z)}, \quad \tilde \psi^{\pr} = \tilde \psi e^{ w (z)}, \quad
R^{\pr} = R + \bar w + w, \quad \varphi_l^{\pr} = \varphi_l
\label{4.3}
\er
and leads to the Noether currents (chiral)
\br
J_{\l_1\cdot H} &=& Tr \(\l_1 \cdot H B^{-1}\pa B \) =  \pa R - {{n+1}\o {n}}\tilde \psi \pa \tilde \chi e^{R-\varphi_2},
\quad \bar J_{\l_1\cdot H} = 0 \nonu \\
\bar J_{\l_1\cdot H} &=& Tr \(\l_1 \cdot H \bar \pa B B^{-1}\) =  \bar \pa R - 
{{n+1}\o {n}}\tilde \chi \bar \pa \tilde \psi e^{R-\varphi_2},
\quad \pa \bar J_{\l_1\cdot H} = 0
\label{4.4}
\er
In addition the model in the group  (\ref{4.1}) has the following topological currents (non chiral)
\br
J_{top, ax}^{\mu} &=& \eps^{\mu \nu } \pa_{\nu} R, \nonu \\
J_{top, vec}^{\mu} &=& \eps^{\mu \nu } \pa_{\nu} u, \quad \quad u= {1\o 2} ln \( {{\tilde \psi }\o {\tilde \chi }}\)
\label{4.5}
\er
When reducing the model from the group $G_0$ to
the coset $G_0/G_0^0$, the chiral symmetry  reduces to 
a global symmetry.   For axial factor group element $g_{0, ax}^{f}$  defined in (\ref{3.2}), has  no 
$\lie_0^0$ generator (i.e., $R$ field to absorb the factor $\bar w + w$ ) 
and  the symmetry of the axial model consists of transformations (\ref{4.3}) such that  $\bar w + w =0$.
For the vector factor group element  $g_{0, vec}^{f}$ in (\ref{3.4}) the
invariance of the field  $t^{\pr} = t$ also implies $\bar w + w =0$. 
In both cases, the remaining symmetry is obtained for $\bar w = -  w = const.$ 
corresponding to a global $U(1)$.
  Further, the reduction from the  group $G_0$ to
the coset $G_0/G_0^0$ is implemented by imposing the subsidiary constraint 
\br
J_{\l_1\cdot H} = \bar J_{\l_1\cdot H} = 0
\label{4.6}
\er 
 which allows the elimination of one degree of freedom.
For the axial gauging we solve the subsidiary constraints (\ref{4.6}) for the nonlocal field $R$, i.e.
the coset $G_0/G_0^0$ is implemented by imposing the subsidiary constraint 
\br
\pa R = \( {{n+1}\o {n}}\) {{\psi \pa \chi}\o {\Delta }}e^{-\varphi_2}, \quad \quad 
\bar \pa R = \( {{n+1}\o {n}}\) {{\chi \bar \pa \psi}\o {\Delta }}e^{-\varphi_2}
\label{4.7}
\er 
where $ \psi = \tilde \psi e^{{1\o 2}R}, \chi = \tilde \chi e^{{1\o 2}R}$ and 
$\Delta = 1+ {{n+1}\o {2n}}\psi \chi e^{-\varphi_2}$.
For the vector gauging, we define $h = 1+\tilde \psi \tilde \chi
e^{-\varphi_2}$ and
 solve the same constraints (\ref{4.6}) for the nonlocal field $u$,
\br
\pa u &=& -{{1\o 2}}\({{n-1}\o {n+1}} +h \) {{\pa R}\o {h-1}} + {1\o 2} 
{{e^{-\varphi_2}}\o {h-1}}\pa \( (h-1)e^{\varphi_2}\), \nonu \\
\bar \pa u &=& {1\o 2}\({{n-1}\o {n+1}} +h \) {{\bar \pa R}\o {h-1}} - 
{1\o 2}{{e^{-\varphi_2}}\o {h-1}}\bar \pa \( (h-1)e^{\varphi_2}\)
\label{4.8}
\er

Finally let us  discuss the relation among the topological and Noether charges of the $G_0$ and the $G_0/G_0^0$.
 Consider the topological charges defined from (\ref{4.5})
\br
Q_{top, ax}^{G_0} = \int_{-\infty}^{\infty}dx  \pa_{x} R = 
\int_{-\infty}^{\infty}dx  \( {{\psi \pa \chi - \chi \bar \pa \psi }\o {\Delta}}\) = 
Q_{Noether, ax}^{G_0/G_0^0},
\label{4.9}
\er
\br
Q_{top, vec}^{G_0} = \int_{-\infty}^{\infty}dx  \pa_{x} u &= &
\int_{-\infty}^{\infty}dx  \(  -\({{n-1}\o {n+1}} +h\) {{\pa_{t} R}\o {h-1}} 
+ {{e^{-\varphi_2}}\o {h-1}}\pa_{t} \( (h-1)e^{\varphi_2}\)   \) \nonu \\
&=& Q_{Noether, vec}^{G_0/G_0^0}
\label{4.10}
\er
by virtue of eqns. (\ref{4.7}) and (\ref{4.8}).  The RHS of eqns. (\ref{4.9}) and (\ref{4.10}) coincide precisely with the
Noether charges associated to the global transformations $\d \psi = \a \psi, \;\; \d \chi = -\a \chi$ and 
$\d \phi_1 = {{(n+1)}\o n} \d R = \a$, $\a =$ const. respectively.

\section{Axial-Vector Duality}

In this section we shall prove that the axial and vector models are 
related by a canonical transformation.
Consider the $SL(3)$  vector model 
\br
{\cal L}_{vec} = \pa ln C \bar \pa ln C - 
{1\o 2} {{\pa A \bar \pa B + \bar \pa A \pa B }\o {1- AB}} - AC^2 + {{B}\o {C^2}}
\label{5.1}
\er
  In terms of  the new set of more convenient variables 
\br
a= AB, \quad C =dA^{-{1\o 2}}, \quad d= e^{{1\o 2}f}, \quad \theta = ln A
\label{5.2}
\er
the lagrangian (\ref{5.1}) becomes
\br
{\cal L}_{vec} =\({{1+3a}\o {1-a}}\) \pa \theta \bar \pa \theta  
-{1\o 4} \pa \theta \(\bar \pa f + {{2\bar \pa a}\o {1-a}}\) 
-{1\o 4} \bar \pa \theta  \(\pa f + {{2\pa a}\o {1-a}}\) + {1\o 4}\pa f \bar \pa f - V_{vec} \nonu \\
\label{5.3}
\er
where $V_{vec} = d^2 + {{a}\o {d^2}}$.  The canonical momenta are given by
\br
\Pi_{\theta} &=& {{\d {\cal L}}\o {\d \dot {\theta}}} = {1\o 2} {{(1+3a)}\o {1-a}}\dot {\theta} - {1\o 2} \( \dot f +
{{2\dot a}\o {1-a}} \)\nonu \\
\Pi_{f} &=&{{\d {\cal L}}\o {\d \dot {f}}} = -{1\o 2} (\dot {\theta} + \dot f )\nonu \\
 \Pi_{a} &=&{{\d {\cal L}}\o {\d \dot {a}}} =- {{\dot \theta }\o {1-a}}
\label{5.4}
\er
and the hamiltonian is 
\br
{\cal H}_{vec} &=& -(1-a)\Pi_{a} \Pi_{\theta} + \Pi_f^2 - (1-a)\Pi_a \Pi_f - a(1-a) \Pi_a^2 \nonu \\
&+& {1\o 4} {{(1+3a)}\o {1-a}} {{\theta}^{\pr}}^2 -{1\o 2}( f^{\pr} +{{2a^{\pr}}\o {1-a}}){\theta}^{\pr} - {1\o 4}
{f^{\pr}}^2  + V_{vec}
\label{5.5}
\er

Consider the following modified lagrangian 
\br
{\cal L}_{mod} ={\cal L}_{vec} - \tilde \theta \(\pa \bar P - \bar \pa P \)
\label{5.6}
\er
where we identify $\pa \theta = P, \quad \bar \pa \theta = \bar P$ \cite{buscher}, \cite{kiritsis}. 
 Integrating by parts and negleting total derivatives,
\br
{\cal L}_{mod} &=&\({{1+3a}\o {1-a}}\)  P \bar P  
-{1\o 4} P  \(\bar \pa f + {{2\bar \pa a}\o {1-a}} + \bar \pa \tilde \theta \) \nonu \\
&-&{1\o 4} \bar P   \( \pa f + {{2\pa a}\o {1-a}} - \pa \tilde \theta \) + {1\o 4}\pa f \bar \pa f - V_{vec} 
\label{5.7}
\er
Integrating over the auxiliary fields $P$ and $\bar P$ we find the effective action
\br
{\cal L}_{eff} ={1\o 4}\pa f \bar \pa f - 
{1\o 4}{{(1-a)}\o {1+3a}} \(\bar \pa f + {{2\bar \pa a}\o {1-a}} + \bar \pa \tilde \theta \)
\( \pa f + {{2 \pa a}\o {1-a}} -  \pa \tilde \theta \) -V \nonu  \\
\label{5.8}
\er
and the canonical momenta are
\br
\Pi_{f} &=&{{\d {\cal L}_{eff}}\o {\d \dot {f}}} = {{2a\dot f}\o {1+3a}} - {{\dot a}\o {1+3a}} + {1\o 2}{{(1-a)}\o
{1+3a}}\theta^{\pr} \nonu \\
\Pi_{\tilde \theta} &=& {{\d {\cal L}_{eff}}\o {\d \dot {\tilde \theta}}} = 
{1\o 2}{{(1-a)}\o {1+3a}} (\dot {\tilde \theta} -f^{\pr} )
- {{g^{\pr}\o {1+3a}}} \nonu \\
 \Pi_{a} &=&{{\d {\cal L}_{eff}}\o {\d \dot {a}}} = {{(\tilde \theta^{\pr} -\dot f)}\o {1+3a}} -{{2a}\o {(1-a)(1+3a)}}
\label{5.9}
\er
The hamiltonian is 
\br
{\cal H}_{mod} &=& -{1\o 2}(1-a)\Pi_{a} {\tilde \theta}^{\pr} + \Pi_f^2 - (1-a)\Pi_a \Pi_f - a(1-a) \Pi_a^2 \nonu \\
&+&  {{(1+3a)}\o {1-a}} {\Pi_{\tilde \theta}}^2 -{1\o 2}( f^{\pr} +{{2a^{\pr}}\o {1-a}})\Pi_{\tilde \theta} - {1\o 4}
{f^{\pr}}^2  + V_{vec}
\label{5.10}
\er
The canonical transformation 
\br
\Pi_{\theta} = -{1\o 2} \tilde \theta^{\pr}, \quad \quad \theta ^{\pr} = -2 \Pi_{\tilde \theta}
\label{5.11}
\er
preserves the Poisson bracket structure and provide the 
 equality of the hamiltanians ${\cal H}_{mod} = {\cal H}_{vec}$. 
If we now substitute
\br
\tilde \theta = 2 ln \( {{\psi}\o {\chi}}\), \quad a = 1+ \psi \chi e^{-\varphi_2}, \quad f = - \varphi_2
\label{5.12}
\er
in the effective  lagrangian $ {\cal L}_{eff}$ (\ref{5.8}), we find
\br
{\cal L}_{eff} = {1\o 2}\pa \varphi_2 \bar \pa \varphi_2 + {{\pa \chi \bar \pa \psi }\o {\Delta }}e^{-\varphi_2} - V
\label{5.13}
\er
which is precisely the axial lagrangian for $\lie_0 = SL(3)$.  Therefore 
the axial and the vector models are related by  the
canonical transformation (\ref{5.11}) preserving the hamiltonians.  In fact,  $\theta$ and $\tilde \theta$ are
isometric variables  generating Noether charges (\ref{4.9}) and (\ref{4.10}) under global transformations.

\section{Concluding Remarks}
We have seen that the crucial ingredient which allows the construction of the axial 
and vector models is the existence of non trivial subgroup $G_0^0$ which in the example 
developed here,  $G_0^0 = U(1)$.  The same strategy works
equally well for  generalized  multicharged NA Toda models  such 
as those constructed in ref. \cite{multi} where $G_0^0 = U(1)\otimes U(1)$ or
 to the homogeneous sine Gordon (HSG) models proposed in
\cite{miramontes}.

The question of constructing integrable models with non abelian internal 
symmetry  arises naturally. For instance, in the case  of  
example $(2c)$ of section 2 (eqn. (\ref{2.11})), 
the group model (\ref{2.5}) admits invariance under chiral 
$\lie_0^0 = SL(2)\otimes U(1)$ transformations and is the natural prototype to describe
 solitons carrying  nonabelian  degrees of
freedom.

An interesting and intriguing subclass of NA Toda models correspond to 
the following three affine  Kac-Moody algebras, 
$B_n^{(1)}, A_{2n}^{(2)}$ and $D_{n+1}^{(2)}$.  Their axial and vector
actions were constructed in \cite{dual} and shown to be identical.  
In fact, those affine algebras satisfy the {\it no torsion condition} 
proposed in \cite{dual} which is fulfilled by  Lie algebras possesseing 
$B_n$-tail like Dynkin diagrams (see for instance appendix N of \cite{cornwell}).
 The very same  selfdual models were shown to possess an exact S-matrix  
 coinciding with certain Thirring models coupled to
 affine abelian Toda models in ref. \cite{Fat}.

{\bf Acknowledgments}  We thank O. Babelon, J.L. Miramontes and
J. Sanchez-Guillen   for discussions.
We are grateful to CAPES, CNPq, FAPESP and UNESP for 
financial support.


\begin{thebibliography}{99}
\bibitem{fadeev}  L.D. Faddeev and L. Takhtajan, ``Hamiltonian methods in 
the Theory of
Solitons'', (Springer, Berlin ),1987

\bibitem{lez-sav}  A. N. Leznov, M. V. Saveliev, Group Theoretical Methods for 
Integration of
Nonlinear Dynamical Systems, Progress in Physics, Vol. 15 (1992), Birkhauser
Verlag, Berlin

\bibitem{ora}
J. Balog, L. Feher, L. O'Raifeartaigh, P. Forgas and A. Wipf, Ann. of Phys. 
{\bf 203} (1990) 76

\bibitem{Aratyn} H. Aratyn, L.A. Ferreira, J.F. Gomes and A.H. Zimerman,
Phys. Lett {\bf B 254} (1991) 372

\bibitem{schw}L.A. Ferreira, J.F. Gomes, A. Schwimmer and A.H. Zimerman, \PLB{274}{1992}{65}




\bibitem{tau} J.F. Gomes, E.P. Gueuvoghlanian, G.M. Sotkov and A.H.
Zimerman, \NPB{598}{2001}{615} , hepth/0011187


\bibitem{dual} J.F. Gomes, E.P. Gueuvoghlanian, G.M. Sotkov and A.H.
Zimerman, \AoP{289}{2001}{232} , hepth/0007116

\bibitem{elek} J.F. Gomes, E.P. Gueuvoghlanian, G.M. Sotkov and A.H.
Zimerman, \NPB{606}{2001}{441} , hepth/0007169

\bibitem{buscher}T. Busher, \PLB{159}{1985}{127},\PLB{194}{1987}{59}, \PLB{201}{1988}{466}

\bibitem{kiritsis}E. Kiritsis, \MPLA{6}{1991}{2871}

\bibitem{multi} I. Cabrera-Carnero, J.F. Gomes, G.M. Sotkov and A.H.
Zimerman, \NPB{634}{2002}{433} , hepth/0201047



\bibitem{cornwell} J.F. Cornwell, Group Theory in Physics, Vol. 3,
Academic Press (1989)


 \bibitem{miramontes} C.R. Fernandez-Pousa, M.V. Gallas, T.J. Hollowood and J.L. Miramontes, \NPB{484}{1997}{609}, 
 \NPB{499}{1997}{673}
 
 \bibitem{Fat}  V.A. Fateev, Nucl. Phys. {\bf B479} (1996) 594
 
 
 
 
 
 
 
 
 
 
\end{thebibliography}
\end{document}